\documentclass[aps,pre,twocolumn]{revtex4-2}

\usepackage{epsf}
\usepackage{latexsym}
\usepackage{epsfig}
\usepackage{amsmath}
\usepackage{amsfonts}
\usepackage{amssymb}
\usepackage{graphicx}
\usepackage{color}
\usepackage{enumerate}
\usepackage{array}
\usepackage[usenames,dvipsnames]{xcolor}

\newcommand{\prs}[1]{{\left(#1\right)}}
\newcommand{\col}[1]{{\left[#1\right]}}
\newcommand{\chs}[1]{{\left\{#1\right\}}}
\newcommand{\avg}[2]{{\left<#1\right>_{#2}}}
\newcommand{\prob}[1]{{\mathcal{P}\prs{#1}}}

\newcommand{\cut}[1]{}

\begin{document}
	
% ================
% Title and Author
% ================
\title{Explaining the Machine Learning Solution of the Ising Model}
\author{Alamino, R.C.}
\affiliation{Aston Centre for AI Research \& Application (ACAIRA), Aston University, Birmingham, B4 7ET, UK}

% ========
% Abstract
% ========
\begin{abstract}
As powerful as machine learning (ML) techniques are in solving problems involving data with large dimensionality, explaining the results from the fitted parameters remains a challenging task of utmost importance, especially in physics applications. This work shows how this can be accomplished for the ferromagnetic Ising model, the main target of several ML studies in statistical physics. Here it is demonstrated that the successful unsupervised identification of the phases and order parameter by principal component analysis, a common method in those studies, detects that the magnetization per spin has its greatest variation with the temperature, the actual control parameter of the phase transition. Then, by using a neural network (NN) without hidden layers (the simplest possible) and informed by the symmetry of the Hamiltonian, an explanation is provided for the strategy used in finding the supervised learning solution for the critical temperature of the model's continuous phase transition. This allows the prediction of the minimal extension of the NN to solve the problem when the symmetry is not known, which becomes also explainable. These results pave the way to a physics-informed explainable generalized framework, enabling the extraction of physical laws and principles from the parameters of the models.
\end{abstract}

\pacs{}
\maketitle
	
% ============
% Introduction
% ============
\section{Introduction}
	
From the second part of the 2010's, the field of condensed matter has experienced an explosion in applications of machine learning (ML) techniques to address phenomena related to phase transitions, with a particular focus on the characterization of phase diagrams for both classical and quantum models  \cite{Wang16,Carrasquilla17,Hu17,Schindler17,VanNieuwenburg17,Chng18,Canabarro19,Zhang20,Huang21a,Yevick22,Baul23}. While the studied models are generally those describing condensed matter systems, phase transitions are also important in the study of complex systems in several disciplines, from biology to social sciences, as they allow the understanding of emergent collective behavior, the hallmark of complexity \cite{Sole11}. 
	
At its heart, ML is statistical inference assisted by computational methods in order to fit parametric functions with a very large number of parameters using also very large datasets. Modern ML models, which include neural networks (NNs), support vector machines and random forests among others, are sophisticated parametric families of functions supplemented by learning algorithms designed to fit the parameters according to appropriate cost functions while avoiding overfitting. 
	
Due to its simplicity, the presence of a continuous phase transition and wide applicability outside physics \cite{Nguyen17}, the two-dimensional classical Ising model -- in particular its ferromagnetic version -- has been one of the main targets of ML applications in the field. It has been demonstrated that it is possible to detect the existence of its two phases (ferromagnetic and paramagnetic) using unsupervised learning \cite{Wang16,Wetzel17} and to approximate the critical temperature for the transition with both unsupervised \cite{Wetzel17b,Liu18,Yevick22} and supervised learning \cite{Carrasquilla17} using as data only the spin configurations at different temperatures. It was also observed that a model trained for a particular lattice topology can infer with success the critical temperature for other lattices, what is usually known in the ML field as transfer learning \cite{Carrasquilla17}, particularly if they belong to the same universality class. 
	
The importance of these results stems from the fact that, although simplified, the described scenario corresponds to the actual discovery and characterization of full phase diagrams purely from microscopic experimental data, providing key information for discovery of new materials \cite{Jansen15, Kusne20, Peterson21}. There is mounting evidence, although there is no general methodology or proofs, that these results can be extended to general phase transitions \cite{Carrasquilla17,VanNieuwenburg17,Zhang17,Lee19,Zhang22}.
	
For the particular case of the ferromagnetic Ising model, increasingly sophisticated methods have been used to tackle the supervised learning problem of classifying configurations according to their phases, from one-hidden-layer NNs (1HLNNs)  \cite{Carrasquilla17,Kim17} to deep learning architectures \cite{Walker20}, such as convolutional neural networks (CNNs) \cite{Wetzel17b} and autoencoders (plain and variational) \cite{Wetzel17, Yevick22}. Although powerful, a drawback of sophisticated state-of-the-art ML models is that the large number of parameters, allied to the varied choices for hyperparameters (number of layers, units per layer, pooling layers, loss functions and others), make it difficult to extract meaning from the trained models -- the more intricate the architecture, the more of a black-box they become. Addressing this point is known as the issue of `explainable AI' (XAI) \cite{Murdoch19}, which became an area of study itself.
	
On the other hand, the unsupervised problem of discovering phases without the prior knowledge of how many or where they are localized on the phase diagram, although apparently more difficult, has been shown to provide the correct results by using principal component analysis (PCA) \cite{Wang16}, one of the simplest methods for feature selection based on linear transformations. The obtained solution detects the direction of the greatest variance of the data as that given by the magnetization per spin of the configurations. Intuitively, the inter-class variance should be larger than the intra-class one if the classes are correctly identified as the phases, justifying the result. Still, in general, there is no further analysis of it. 
	
Being able to obtain insights on how a trained ML model solves a problem is fundamental for breaking barriers against their general use, a known problem in health applications where distrust and ethical concerns by professionals who are not ML specialists (as well as adverse legal implications) may lead to decreased adoption, preventing the area from reaping the potential benefits \cite{Rajpurkar22}. While the adoption barrier might be lower in areas with less direct human interaction, explainability has important technical and scientific advantages. Unveiling the mechanisms used by a model to infer the necessary patterns directly leads to improvements in its architecture and, consequently, to clearer and more efficient solutions. A yet more fundamental reason for seeking it goes to the heart of physics and scientific research itself. Hidden patterns encoded in the structure of the trained models might reveal new physical laws, such as symmetries, leading to new discoveries and further understanding. 
	
The aim of this work is to carry out such an analysis. While previous authors \cite{Carrasquilla17, Wetzel17b} have taken important steps towards understanding ML solutions, being able to identify several aspects of the obtained inference, here new results are presented. For the unsupervised learning case, it is demonstrated that the PCA solution clearly detects that the greatest difference in the order parameter happens when the temperature is varied by deriving a measure of variance over the whole temperature interval. 

While a whole set of techniques has been developed for XAI, which include notorious algorithms like SHAP \cite{Lundberg17} and LIME \cite{Ribeiro16}, several of them focus on detecting which input features to the model are most influential in the obtained predictions. For the Ising model, the features are the spins, which should all contribute equally to the prediction and, therefore precluding the applicability of the such analyses.  
	
For the supervised learning case, it is worth pointing out a key result from the work by Kim and Kim \cite{Kim17}, who were able to predict and prove that a 1HLNN needs only two hidden units to identify the critical temperature. By analyzing this simplified model, they concluded that the NN used the spin inversion symmetry of the Hamiltonian at zero external field to learn the scaling dimension of the magnetization. 
	
The present work goes even further to find a yet simpler NN (arguably the simplest) that can pinpoint the critical temperature of the two-dimensional ferromagnetic Ising model, which from now on will be referred just as `the Ising model' for short. Complementing the work of Kim and Kim \cite{Kim17}, a full explanation of the solution can then be found, leading to a better understanding of the success of previous ML approaches. 
	
The technique used here is at the core of physics research -- finding the simplest non-trivial model that is capable of solving the task allows a clearer vision of its inner workings. The solution is found using a NN without any hidden layers, which will be called a single-layer NN (SLNN) for short. Such an architecture is also widely known as Rosenblatt's perceptron \cite{Rosenblatt62} or the McCulloch-Pitts neuron \cite{McCulloch43}. The SLNN is trained with data from a single-sized square lattice, without the need for regularization and with data excluding a relatively large asymmetric interval around the known critical temperature. This is accomplished by a physics-informed approach, although not in the same sense as in the work of Karniadakis \textit{et al.} \cite{Karniadakis21}, where physical information is included in the model's loss function. Here, minimal knowledge of the physics of the problem is used to justify the choice of the architecture, in this case the symmetry of the Hamiltonian by full spin inversion. The trained network cannot only find very close estimates for the square lattice, but also for other two-dimensional lattices and, although with less precision, for the cubic one -- which is expected as the latter belongs to a different universality class. 
	
To set the notation, let us write the Hamiltonian of the Ising model as
\begin{equation}
	H = -\sum_{\avg{i,j}{}} J_{ij}s_is_j,
	\label{equation:HIM}
\end{equation}
where $s_i\in\chs{-1,1}$ for $i = 1,..., N$, with $N$ the number of spins in the system and $J_{ij}$ the exchange couplings. Local external magnetic fields are considered to be zero in order for the critical phase transition to occur at a finite temperature. For each configuration $s = (s_1, s_2,...,s_N)$ of the lattice, the magnetization per spin is given by the average of all the spin values on it $m(s) =\sum_i s_i/N$, i.e., an equal linear combination of them. Throughout this work, the Boltzmann constant is set to unit, $k_B=1$. 
	
Onsager's solution for an infinite square lattice \cite{Onsager44} assumes the couplings $J_{ij}$ are the same in each of the two perpendicular directions in the lattice, calling them $J_1$ and $J_2$. Here, let us assume for simplicity that $J_1=J_2=J$. The ferromagnetic transition requires $J>0$, which is the situation considered throughout this work. The critical temperature is given by $T_c = 2J/\ln\prs{1+\sqrt{2}}$ and the average magnetization per spin at temperature $T$ is $\avg{m}{T} = \pm \prs{1-\sinh^{-4} 2\beta J}^{1/8}$, $0\leq|\avg{m}{T}|\leq1$, where $\beta = 1/T$ and the sign is due to the $\mathbb{Z}_2$ symmetry of the ferromagnetic phase. For $T<T_c$, $|\avg{m}{T}|\neq0$ and the system is in the ferromagnetic ordered phase, while for $T>T_c$, $\avg{m}{T}=0$ and the system is in the paramagnetic disordered phase. The value of the average magnetization completely identifies the phase (ordered/disordered) of the system at a certain temperature in the limit of an infinite lattice.
	
\section{Unsupervised Phase Detection}
	
The problem of unsupervised detection of phases for the Ising model consists of identifying to which phase a given configuration of spins most probably belongs without providing to the ML model the number of possible phases or their boundaries on the phase diagram. All the information has to be inferred directly from the data. Such problems are usually formulated as clustering problems, i.e., clustering datapoints within the same class if they have similar characteristics, the latter being dependent on the algorithm that is used. 
	
Strictly speaking, PCA is not a clustering method, but what is called a feature or dimensionality reduction method. It is based on the idea that, if the datapoints to be analyzed are linearly correlated, one can find a direction in the feature space (features corresponding to the coordinates of such points) in which the variance assumes its greatest value \cite{Bishop06} -- the so-called first principal component. The rational is that, inside the same class, the relevant features should not vary too much, otherwise the very concept of a class of similar datapoints would not be sensible. That was the technique first used by Wang \cite{Wang16} to identify the phases of the Ising model and to detect the order parameter as the magnetization. There, the author shows that, when Ising configurations are projected into the directions of the first two principal components, the separation into clusters becomes clear enough to be obtained by simple clustering method ($K$-means, in this case), with the main direction of separation being given by the first component.     
	
For the Ising model, the PCA solution is obtained by first generating the configurations of the model often by Monte Carlo (MC) methods \cite{Landau05}. The dataset is composed by generating a total of $M$ configurations of $N$ spins $s^\mu \in \chs{\pm1}^N$, $\mu=1,...,M$, for $K$ different values of temperature $T_k$, $k=1,2,...,K$. For each particular temperature, $M_k$ configurations are generated, with $\sum_{k=1}^K M_k=M$. The covariance matrix $C$ of the dataset is given by
\begin{equation}
	C = \frac1M \sum_{\mu=1}^M (s^\mu-\bar{s})(s^\mu-\bar{s})^T,
\end{equation}  
where the $s^\mu$ are understood as column vectors and 
\begin{equation}
	\bar{s} = \frac1M \sum_{\mu=1}^M s^\mu,
\end{equation}
is the mean configuration of the whole dataset.
  	
PCA is implemented by finding the direction $u$ in configuration space in which the projection of $C$ is maximized. The solution is obtained by maximizing the quantity
\begin{equation}
	\mathcal{L} = u\cdot Cu +\lambda(u\cdot u-1),
\end{equation}
where $\lambda$ is a Lagrange multiplier and the last term guarantees that $u$ has norm 1. The solution is simply $Cu=\lambda u$, which means that $u$ is the eigenvector of $C$ with the largest eigenvalue $\lambda= u\cdot Cu$.
 		
The above solution can be easily obtained numerically and, if all $M_k$ are the same, it approximates $u=\frac1{\sqrt{N}}(1,1...,1)$ with high precision. The projection of any normalized configuration into this direction, provides its magnetization per spin, which is the order parameter of the model. Using the same amount of datapoints for each temperature guarantees that the results are not biased towards any of the phases and is essential to identify the correct order parameter.

The obtained result is therefore interpreted as PCA identifying the order parameter of the phase transition. The projections of the configurations along this direction form three clusters corresponding to the two ferromagnetic phases, one with positive and the other with negative magnetization, and the paramagnetic phase (see \cite{Wang16} for a visual representation). 

Let us now show that this particular choice corresponds, in fact, to the model identifying that the gradient of the magnetization points towards the direction of temperature gradient, a result that has not been pointed out before. In order to see that, let us calculate the variance $u\cdot C u$ in the direction of the magnetization eigenvector
\begin{equation}
	\begin{split}
		\sigma^2_u 	&= \frac1N \sum_{i,j} C_{ij} = \frac1N \sum_{i,j} \frac1M \sum_{\mu=1}^M \prs{s_i^\mu - \bar{s_i}}\prs{s_j^\mu - \bar{s_j}}\\ 
				    &= \frac1M \sum_{\mu=1}^M\col{\frac1{\sqrt{N}} \sum_i \prs{s_i^\mu - \bar{s_i}} }^2\\
					&= \frac1M \sum_\mu\col{\frac1{\sqrt{N}} \sum_i \prs{s_i^\mu - \frac1M \sum_\nu s_i^\nu} }^2\\
					&= \frac{N}{M} \sum_\mu\prs{m^\mu - \frac1M \sum_\nu m^\nu}^2,
	\end{split}
\end{equation}
which implies
\begin{equation}
	\frac{\sigma^2_u}N = \frac1M \sum_\mu\prs{m^\mu}^2 - \prs{\frac1M \sum_\mu m^\mu }^2,
\end{equation}
where $m^\mu$ is the magnetization of the $\mu$-th configuration in the dataset. Define the average magnetization at temperature $T_k$ as
\begin{equation}
	\avg{m}{k} \equiv \frac1{M_k} \sum_{\mu|T(\mu)=T_k}  m^\mu,
\end{equation}
where the subscript of the sum means all configurations $\mu$ which were generated at the same temperature $T_k$, and analogously for the second moment. Then, one can rewrite the variance per spin as
\begin{equation}
	\frac{\sigma^2_u}N = \sum_k w_k \avg{m^2}{k} - \prs{\sum_k w_k \avg{m}{k}}^2,
\end{equation}
where $w_k = M_k/M$ is the frequency of a certain temperature in the measurements. 

As mentioned before, the direction of the magnetization coincide with the first principal component when $M_k=M/K$ for all $k$, the choice that leads to the discussed PCA result. Then, we can write
\begin{equation}
	\frac{\sigma^2_u}N  = \sigma^2_T +\avg{\sigma^2_k}{}, 
\end{equation}
where
\begin{align}
	\sigma^2_T &= \frac1K\sum_k\prs{\avg{m^2}{k}-\avg{m}{k}^2},\\
	\sigma^2_k &= \frac1K\sum_k\avg{m}{k}^2 -\prs{\frac1K\sum_k\avg{m}{k}}^2 = T_k \chi_k,
\end{align}
and $\avg{\sigma^2_k}{}=(1/K)\sum_k \sigma^2_k$ with $\chi_k$ being the magnetic susceptibility at temperature $T_k$. The quantity $\sigma^2_T$ measures the total fluctuation of the average magnetization as the temperature is varied, i.e.,  in the direction of the temperature gradient. This is going to dominate over $\avg{\sigma^2_k}{}$, which measures the fluctuations of the magnetization at each temperature. This allows PCA not only to pick up the order parameter, but at the same time identify the temperature as the relevant control parameter of the phase transition.   

As we will show later, the success in finding the appropriate phases of the Ising model combining PCA with $K$-means clustering as described in \cite{Wang16} will be a direct result of the linear separability of the phases in configuration space when the sign of the magnetization is constrained to either positive or negative (but not both). More sophisticated, non-linear clustering algorithms are then unnecessary in the case of the Ising model, although they might be needed for more general models with a more complex phase diagram.
 	
\section{Supervised Detection of the Critical Temperature}

The supervised learning problem of detecting the critical temperature using ML methods can be stated as follows. In the same way as in the PCA case, assume that a dataset composed of spin configurations of the system at different temperatures is provided for the two-dimensional ferromagnetic Ising model on a square lattice defined by the Hamiltonian from equation (\ref{equation:HIM}). Periodic boundary conditions (PBC) are assumed from now on. Extracting information from this dataset, the task is to find the critical temperature assuming that it is known that there are two phases, one at low and another at high temperatures (although the low temperature phase is in fact two that are clustered together due to the $\mathbb{Z}_2$ symmetry). 

The previously-cited works show that ML methods can solve this problem reasonably well. The ML algorithms need only to be trained with configurations and their respective phases, without the need for providing the temperature in which they were generated. This is due to the fact that the relative probabilities of the configurations in each phase can be inferred by the ML models. Once again, the number of configurations per temperature should be the same for all temperatures in order to allow the model to infer their correct distribution without any bias.

Different strategies for finding $T_c$ have been used in the literature. Here, it is shown that a SLNN -- the simplest possible NN architecture -- can be trained to successfully find the probability of the ferromagnetic phase given a certain configuration $s$. The parameters of the network are its $N$-dimensional weight vector $w$ and its scalar bias $w_0$. The output of the NN is the parameterized probability $\prob{F|s,w, w_0}=\sigma(w\cdot s+w_0)$ of the configuration $s$ belonging to the ferromagnetic phase, where the dot represents the scalar product. The function $\sigma(\cdot)$ is known as the NN's activation function and is usually chosen according to the characteristics of the problem. The SLNN is trained with input/output pairs with the configurations as inputs and the phases as outputs. In ML language, it is trained to solve a binary classification problem using supervised learning.

Because the objective is to fit a probability, the sigmoid function $\sigma(x)=(1+e^{-x})^{-1}$ is chosen here. The fitting of the weights is obtained by minimizing the cross entropy between the predicted distributions of classes and the empirical ones from the training dataset. This choice is equivalent to the common logistic regression, which was known long before NNs were invented, the fitting of which is obtained by maximum likelihood \cite{Bishop06}. Similar results can be obtained by varying the activation function as long as it is chosen sensibly. It is worthwhile noticing that logistic regression has been used to study aspects of the Ising model before using the same input features as here, the configurations, but in the analysis of a different problem \cite{Huang21a}. 

More specifically, the training set is the dataset composed of pairs $D = \chs{\prs{s^\mu, y^\mu}}_{\mu=1}^{|D|}$, where $|D|$ is the number of pairs used for the network's training and $y^\mu =F$ (the ferromagnetic phase) if the equilibrium configuration $s^\mu$ was generated at a temperature $T<T_c$ and $y^\mu = P$ (the paramagnetic phase) if it was generate at $T>T_c$. For simplicity, assume $J=1$. The equilibrium configurations at each temperature are generated using the Metropolis-Hastings (MH) algorithm \cite{Landau05}. Knowledge about the physics of the problem enters at this stage. Considering that the zero field Hamiltonian is symmetric under total spin inversion, without loss of generality the initial state of the Markov chain can be chosen as a lattice with all spins equal to +1. This guarantees that the equilibrium configurations below the critical temperature will have positive magnetization with high probability, simplifying the analysis and decreasing the variance in the statistics of the ferromagnetic states. This choice leads to longer relaxation times for the configurations, but does not prevent pinpointing the correct critical temperature, as it will be demonstrated. 

Previous works used a range of temperatures including $T_c$ to train the model. Here, an extra challenge to the NN is created by restricting the training set to the union of two subintervals of temperature $T\in[0.05, 1]\cup[4,5]$ which, for $J=1$, will not include $T_c\approx 2.269$. The smallest temperature is not zero only for numerical issues (avoiding division by zero) and care is taken not have a symmetric gap around $T_c$ to avoid a false success by simply interpolating linearly the classification. The values of $w$ and $w_0$ are obtained by a gradient descent method. For a large number of parameters, regularization terms are generally needed to avoid overfitting, but tests showed that the results presented here do not change if they are not included.

Once the training is completed, $T_c$ is estimated by generating $L$ test sets composed only by spin configurations, generated also using the MH algorithm. For each of these sets, the probability for each configuration to belong to the ferromagnetic phase is calculated and averaged for each temperature. The estimated $T_c$ for that particular set is given by the first temperature at which the probability drops below $1/2$, i.e., when the model is most confuse about the phase classification on average (known in ML as the decision boundary). A better precision can potentially be obtained by estimating the closest point to the probability 1/2, but the aim of this letter not being a high precision calculation of the critical temperature, the different is not very relevant. The final estimate, with its statistical error, is obtained by considering the estimates for all test sets.

The code used for this work was written in Python. The MH algorithm is standard \cite{Landau05}. The NNs were trained using the Keras library, with the optimization of the cross entropy being carried ou by an Adam optimizer \cite{Kingma14}, which belongs to the class of stochastic gradient descent algorithms. The supervised learning was carried out for 20 epochs of cross validation with a 30/70 split.

Table \ref{table:Tc} shows the different estimates of $T_c$ obtained for the triangular, square, hexagonal and cubic lattices together with the theoretical values for the first three and the numerical estimate for the cubic one. 

\begin{table}[ht]
	\centering
	\begin{tabular}{|c|c|c|}
		\hline
		Lattice    & Known $T_c$ & ML $T_c$          \\
		\hline
		Hexagonal  & 1.518       & 1.538 $\pm$ 0.035 \\
		Square     & 2.269       & 2.260 $\pm$ 0.056 \\
		Triangular & 3.641       & 3.573 $\pm$ 0.072 \\
		Cubic      & 4.511       & 4.311 $\pm$ 0.073 \\
		\hline
	\end{tabular}	
	\caption{Comparison between the known values of the critical temperatures (Known $T_c$) and the ones calculated the SLNN (ML $T_c$). The ML result is the average over 20 realizations of the procedure and the error is the obtained standard deviation. }
	\label{table:Tc}
\end{table}

The results were obtained by training the NN with configurations and phases of a $20\times20$ square lattice in the two different temperature subintervals mentioned before. The SLNN has, therefore, $N+1=401$ parameters. Each of the temperature intervals contains 100 equally spaced temperatures and, for each temperature, $M=1000$ configurations were used for the training after discarding the first 10000 configurations generated by MH in order to reach thermodynamic equilibrium. The order in which the configurations were presented to the network was randomized, although tests show that the results do not change if it is not.

The test sets generated to find the critical temperature have $M'=500$ configurations per temperature for each two-dimensional lattice after discarding the first 1000 configurations. The configurations were generated for 100 temperature points equally spaced in the respective temperature interval, which was extended to include the critical temperatures for each case and has no gaps. Each estimated $T_c$ is the mean over $L=20$ independently generated datasets for each lattice, the latter with sizes $N=400$ for the triangular and square lattices, $N=441$ for the hexagonal and $N=512$ for the cubic. The value of one standard deviation is also provided.

The result for the two-dimensional lattices, including one standard deviation, already encompasses the exact values, which demonstrates the strong ability of such a simple NN to obtain a good approximation for the critical temperatures, even being trained on different intervals and on a different lattice. The known value for the cubic lattice has an error that falls within three standard deviations of the known result, which is not a bad approximation considering that it belongs to a different universality class than that of the two-dimensional lattices as mentioned before. There was no need for any finite-size scaling and, for the hexagonal and cubic lattices, where for practical reasons the actual lattice sizes were greater than 400, the result was obtained using only the first 400 spins of the configurations. Further tests confirm that any 400 hundred spins chosen randomly from the lattices provide the same results, as they should.   

\begin{figure*}[ht]
	\centering
	\includegraphics[width=18cm]{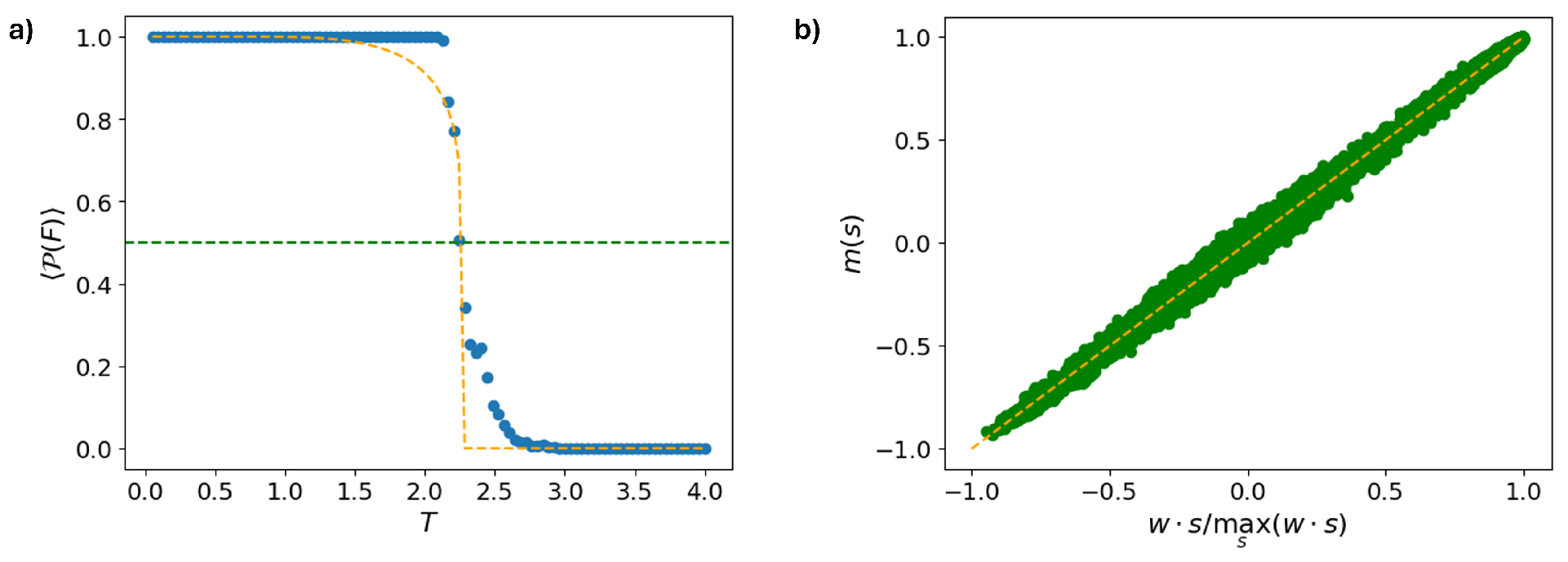}
	\caption{(a) Average probability of being in the ferromagnetic phase averaged over 20 test sets (circles) together with the magnetization per spin (orange dashed line) from Onsager's solution. The horizontal green dashed line marks the decision boundary. (b) Scatter plot showing the almost perfect positive correlation between the magnetization of the spin configurations $s$ and the scaled value of the products $w\cdot s$ for one of the test sets from (a).}
	\label{figure:psig}
\end{figure*}

Let us now focus on analyzing and explaining how the SLNN successfully finds the solution for the problem. Figure \ref{figure:psig}a shows a plot of the average probability $\avg{\prob{F}}{}$ for classifying the configurations as ferromagnetic (circles) for each temperature in the square lattice together with the exact magnetization for the infinite lattice (orange dashed line), averaged over 20 randomly generated test sets.

The first notable feature is that $\avg{\prob{F}}{}$ follows closely the magnetization at the training temperatures and has a transition almost at the same point, even not being trained with information in its direct neighborhood. This is a general feature, with tests showing that changing sensibly the activation function can improve the coincidence, although it does not have a significant effect in the final estimation of $T_c$. In particular, one can see how the point where the probability falls to 1/2 approximates well the critical temperature. That is the criterion used in several works (e.g., \cite{Chng17, Canabarro19}) and, although not explicit in it, the data collapse used to pinpoint the critical temperature in \cite{Carrasquilla17} shows a crossing of the plots that visually coincides with an output equals to 1/2 of the used NN with one hidden layer, which is exactly its decision boundary.  

For a linear classifier, as the SLNN, the decision boundary is the value of $s$ that solves the equation $x \equiv w\cdot s + w_0=0$, where $x$ is the argument of the activation function. Because the symmetry of the lattice with PBCs implies that all sites should contribute equally to the classification, on average all coordinates of $w$ should be equal. This will not happen in practice due to the stochastic nature of the training, but if one considers the asymptotic average scenario, one can predict that the solution should approximate $w_i=\bar{w}$, $\forall i$, leading to $x=\bar{w}Nm+w_0$. 

Figure \ref{figure:psig}b shows a scatter plot of all values of the magnetization $m(s)$ for one single test set (chosen arbitrarily) against the corresponding values of the dot product $w\cdot s$ divided by its maximum value over the set. It is clear from the plot that, although not perfectly, the parameters are indeed fitting the magnetization, apart from a multiplicative constant. 

\begin{figure*}[ht]
	\centering
	\includegraphics[width=14cm]{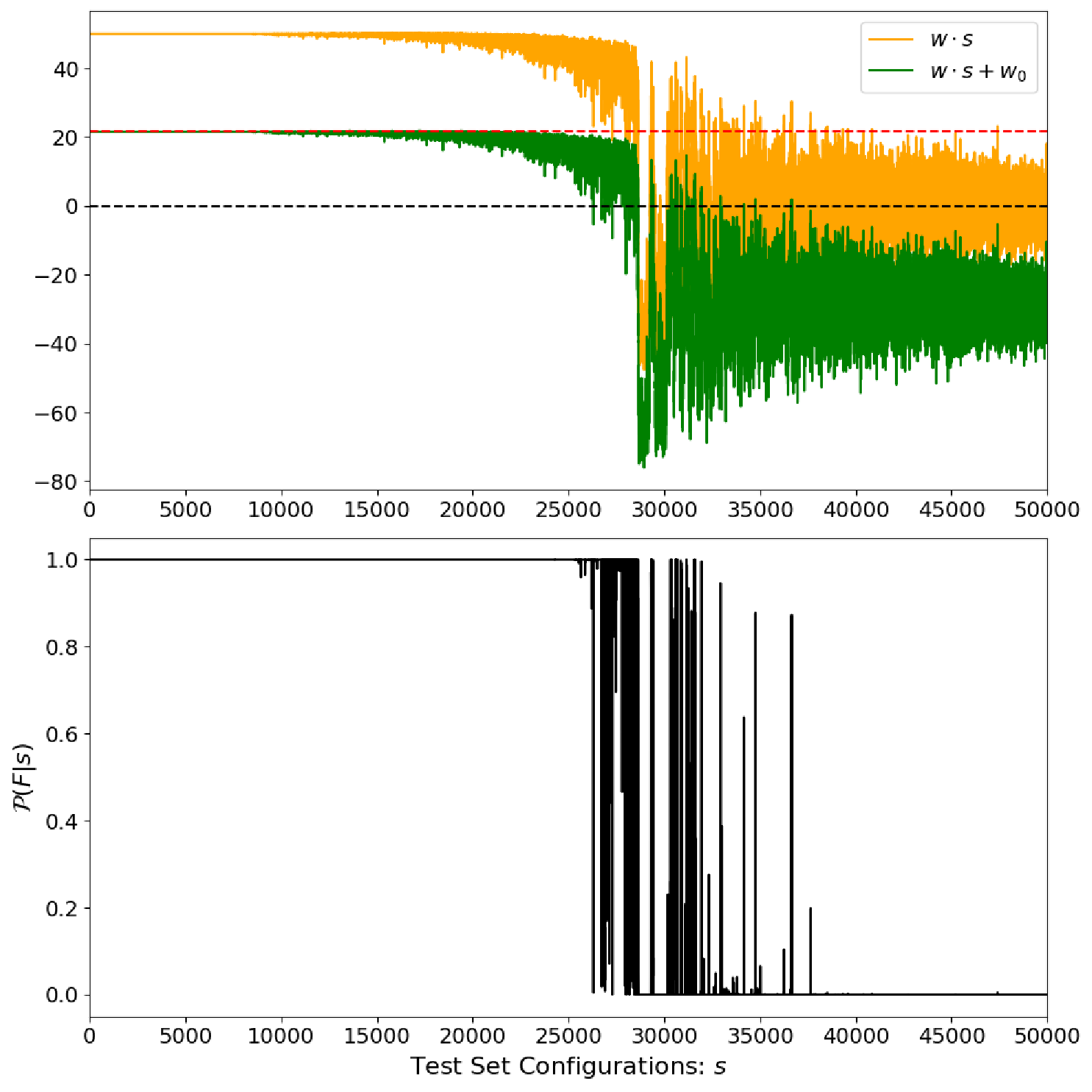}
	\caption{Upper plot: argument of the sigmoid for each configuration in the test set; lower plot: probability of the configuration belonging to the ferromagnetic phase.}
	\label{figure:understand}
\end{figure*}

Figure \ref{figure:understand} reveals yet another piece of information needed to understand the strategy of the model.  The upper plot shows the value of $w\cdot s$ (orange) and that of $x$ (green). The upper (red) dashed line shows the difference between the maximum value of the former and that of the latter. The lower plot is the probability of the corresponding configuration being in the ferromagnetic phase. 

The plots clearly show how the network is working. It first scales up the values of the magnetization by a factor large enough such that the sigmoid's argument $x$ becomes largely negative at low temperatures, bringing the calculated probability of the ferromagnetic phase very close to 1. In order to fit a (nearly) zero probability for the paramagnetic phase, $w_0$ is made negative enough to bring all configurations with magnetization in the training range below zero, but still not enough to spoil the fitting of the ferromagnetic phase. This is achieved, in all observed training, by using a value $w_0$ about half of the maximum $x$. In the region close to the critical temperature, the network simply uses the statistics of the magnetizations learned from the trained intervals. As it depends only on the statistics of the magnetization, the same strategy should allow the network to infer the critical temperature for any lattice -- which indeed is observed to happen. 

The fact that the SLNN is a linear classifier allows for a geometric interpretation of the result. It works by finding the best separating hyperplane between the two classes, in this case the ferro and paramagnetic phases. For the ferromagnetic Ising model, the training rotates the normal to this plane, which is $w$, until it is aligned to the direction of the magnetization -- the same direction found by PCA as the one with maximum variance of the inputs. When trained with positive magnetizations above the critical temperature, as is the case here, the SLNN adjusts the origin of the separating hyperplane such that all configurations in the training set that are in the paramagnetic phase will lie under the plan (in the opposite direction of $w$) having then a probability less than 1/2 of being ferromagnetic. The value of $w\cdot s+w_0$ measures the distance between the configuration and the hyperplane, with the sign indicating in which side of it the configuration lies. If the distance to the plane is large enough, the scaling stretches it by such an amount that it becomes effectively 1 above it and 0 below.

The reason a linear model classifier trained at extreme temperatures in the square lattice can infer the critical temperature for the others, even when only a sample of each configuration is used and the topology is ignored in the inputs is that, as mentioned, the ferromagnetic transition depends only on the value of the magnetization, which is linear and local.

Given the above explanation, it is not a suprise that, if the training is done with the MH algorithm allowing negative magnetizations in the ferromagnetic phase, the SLNN will not be able to learn the classification anymore as one single plane cannot separate the phases in the configuration space. On the axis of the magnetization, the configurations with high probability of being ferromagnetic occupy the regions closer to the values $\pm1$, with the paramagnetic ones located in a cluster between these two. On the other hand, this picture implies that two such planes aligned with the magnetization axis should be enough if their classifications are properly combined. This explains why the 1HLNN with only two units used in \cite{Kim17} is effective in solving the problem. Each unit in a hidden layer works as a separate SLNN with the same input vector, allowing for the segmentation of the space by several planes. By using the understanding obtained with the simpler model, one can then predict the minimum value of a hyperparameter, namely the number of hidden units, necessary to find the correct solution. 

\begin{figure*}[ht]
	\centering
	\includegraphics[width=18cm]{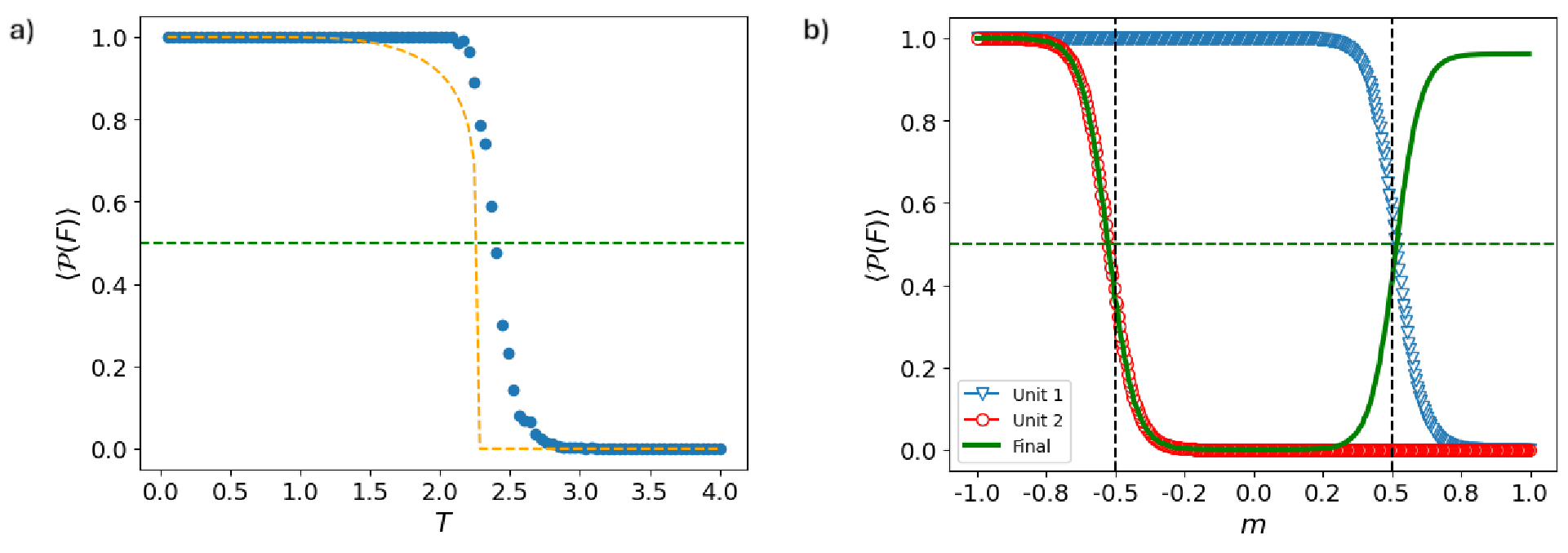}
	\caption{(a) Same plot as in figure \ref{figure:psig}a, but now for the 1HLNN with two hidden units and trained without restrictions on the sign of the magnetization. (b) Probabilities inferred by the units from the hidden layer, named Unit 1 (inverted triangles) and Unit 2 (circles), together with the final probability for the input configuration to be in the ferromagnetic case (thick line) plotted against the magnetization values of the configurations. The plot shows the average over 500 random configurations for each value of the magnetization.}
	\label{figure:apfhl}
\end{figure*}

By analyzing the solution of such a 1HLNN with two units, it is possible, once again, to explain how it is obtained. Both the training and test sets where now generated by MH with ferromagnetic configurations allowed to have negative magnetizations. Figure \ref{figure:apfhl}a shows the average probability of being in the ferromagnetic state over 20 test sets. 

Although less closely, the average probability for the ferromagnetic phase still follows the magnetization for the infinite lattice. The critical temperature is not as well approximated, but although this could be improved \cite{Kim17} by increasing the lattice size and using larger training and test sets, this will not be done here as the aim is to explain the inferred result, not to actually calculate $T_c$. For this particular instance of training, figure \ref{figure:apfhl}b allows us to understand the strategy used by the network to carry out the classification.

The plot is obtained by the following technique. For each fixed value of the magnetization, 500 random configurations are generated. These configurations are then fed to the trained NN and the outputs of each of its units are recorded. The final result is obtained by averaging the outputs and plotting them. On the plot, `Unit 1' and `Unit 2' correspond to the two units of the hidden layer, while 'Final' corresponds to the output unit. The value obtained by Unit 2 is given by the (red) circles shows that this unit specializes in classifying correctly only negative magnetizations, classifying together both zero and positive ones. The separating hyperplane is located close to a magnetization value of -0.5 and magnetizations below that have a probability above 1/2 of being ferromagnetic according to this unit, while those above it are classified as paramagnetic. Of course, this unit completely misclassifies the configurations with positive magnetization. The behavior Unit 1, represented by the (blue) inverted triangles, is more interesting. First, it specializes in recognizing the positive magnetizations. However, it actually exchanges the ferromagnetic by the paramagnetic phase when classifying them. The separating hyperplane is located close to the magnetization value of 0.5, but it gives magnetizations above that a larger probability of being paramagnetic. This ``mistake'' is however corrected by the Final unit in the output layer and the final classification, give by the thick continuous (green) line, provides the correct probability profile for the classification task. 

The exchanged classification of Unit 1 in the hidden layer is not in fact a mistake. Units from the hidden layer have no direct access to the training labels. This information is passed to it only by the output layer through backpropagation. For this layer, labels have only a relative meaning and Unit 1 ends up using the opposite labeling convention from the training set. That is, however, compensated by the output layer which has direct knowledge of the correct labels. It is possible to observe, by repeating the training several times, that all four combinations of label attributions by the hidden units can happen with the same probability, with the final classification always being corrected by the output layer.

The way the output layer combine the probabilities, compensating for any difference in labeling by the hidden units, is deceptively simple. There are only three parameters in this layer, the weights from hidden units 1 and 2 and the bias. For the particular result shown in figure \ref{figure:apfhl}b for instance, these numbers are $w_1\approx-26.691$, $w_2\approx 27.759$and $w_0\approx13.009$, respectively. It is not difficult to convince oneself that the output of the network is obtained by simply rescaling appropriately the combination $-\prob{F}_1 + \prob{F}_2 +1/2$, which is the simplest algebraic expression that will provide a good approximation for the classification. For all observed training sessions, although the exact value of the rescaling changes, the strategy of using this simple algebraic expression is always used by the output layer. 

Simple tests show that the 1HLNN with two units is also capable of finding the critical temperature of the anti-ferromagnetic Ising model using the same strategy when trained with data from that version of the model. The only difference is that the normal to the separating hyperplanes is now in the direction of the staggered magnetization, i.e., the sum of the magnetizations of the two square sublattices with opposite signs. All else remains the same. These results, not shown here, can easily be obtained simply by repeating the whole methodology using $J=-1$ when generating the training and test sets.

% ===========
% Conclusions
% ===========
\section{Conclusions}

There are three main contributions in this work. The first is finding the meaning of the variance in the direction of the first principal component for the PCA application in finding the Ising mode phases using unsupervised learning. The decomposition of this quantity in two terms, one detecting fluctuations in the direction of the temperature gradient and the other being an average of the intra-temperature fluctuations (effectively, the susceptibilities), allowed to see that PCA, in addition to identifying the order parameter itself, also indicates the control variable of the transition.

Secondly, it was shown that, by using the physical knowledge that the Hamiltonian of the Ising model is invariant by total spin inversion, the previous record on the smallest NN that can infer its critical temperature can be broken. This symmetry can be used to choose the architecture of the NN such that it has no hidden layers, which was called a SLNN. This maximally simplified architecture, with the specific choice of a sigmoid as its activation function, is in fact equivalent to plain logistic regression. Finally, it is shown that the architecture of the SLNN allows the full explanation of the solution found by the ML model. Furthermore, this knowledge leads straightforwardly to the prediction that a NN with two units in one single hidden layer can find $T_c$ for the general case in which the magnetization sign is not restricted, explaining in a simpler way previous results.

The explanation of the ML solution for the ferromagnetic Ising model provided here does not invalidate or diminish the importance of any previous results. Instead, it corroborates them by showing that one can indeed understand their inner workings. The explanation of the strategies used by simpler ML models indicates the directions that can be taken in order to fully explain those of the more sophisticated ones, allowing us to start opening the ML black-boxes. In addition, it also indicates the possible paths to address condensed matter systems with phase transitions with more complex phase diagrams where one might find even more phases and more involved order parameters. From the analysis presented here, the presence of more phases clearly requires a corresponding increase in the number of hidden units of the NNs. Depending on the geometry of the corresponding clusters of configurations, kernel techniques \cite{Bishop06} might help decrease the complexity of the required NN architectures in the supervised learning case, while their unsupervised detection will probably require the use of non-linear clustering techniques. Other important cases are the presence of hidden order parameters, which might require extra layers, and general topological phase transitions, which might require the use of CNNs as the lattices would need to be considered as whole ``images''. 

Extensions of the work presented here are currently being applied to the study of other physics systems with more general phase transitions, both classical and quantum. While it has not yet been possible to extract physical laws from the inferred parameters, this remains one of the objectives being pursued by this research.

% ================
% Acknowledgements
% ================
\section*{Acknowledgments}

The author would like to acknowledge insightful discussions with Dr Juan Neirotti, Dr Felipe Campelo and Prof David Saad.

% ============
% Bibliography
% ============
\bibliographystyle{apsrev4-2}
\bibliography{statphysml}
	
\end{document}